\documentstyle[12pt,epsf]{article}
\newcommand{\bm}{\bibitem}

\def\be {\begin{equation}}
\def\ee {\end{equation}}
\def\bea {\begin{eqnarray}}
\def\eea {\end{eqnarray}}

\begin{document}
\normalbaselineskip = 24 true pt
\normalbaselines
\thispagestyle{empty}
\rightline{\large\sf SINP-TNP/97-06}
\rightline{\large May 1997}
\vskip 0.5 cm
\begin{center}{
{\Large {The decay of the omega meson at finite temperature }}
\vskip 1.2 cm
{\bf Indrajit Mitra and Abhee K. Dutt- Mazumder{\footnote
{email: abhee@tnp.saha.ernet.in}}},\\ 
Saha Institute of Nuclear Physics\\
1/AF, Bidhan Nagar\\ Calcutta - 700 064, India
}
\end{center}
\vskip 1.0 cm
\begin{abstract}
The decay width of the $\omega$ meson at finite temperature is
calculated using the Gell-Mann Sharp Wagner model of $\rho$ pole dominance. 
Effective masses of the $\rho$ and $\omega$ are determined within the
framework of real-time formalism of finite temperature field theory.
It is shown that even though the mass of the $\omega$-meson decreases with
temperature, its decay width increases because of an
interesting interplay between the phase space factors, the transition
matrix element and Bose enhancement.
\end{abstract}
\newpage
Properties of nuclear matter at finite temperature have 
recently acquired particular interest as in the 
laboratory such exotic conditions can now be created
by colliding heavy nuclei. In this context, therefore, the behaviour 
of hadrons
at high temperature and/or high density have been investigated by several
authors \cite{song95,gale91,pisarski95}. In particular, efforts have been 
directed towards the unraveling of the properties
of light vector mesons at finite temperature because they show up
in the dilepton spectra observed experimentally. Li, Ko and Brown have  shown
that the dropping vector meson masses can account for the enhanced yield 
of the dileptons in the low mass region as observed in SPS, CERN \cite{li96}. 

Issues such as whether the vector meson masses go up or down with temperature
or how  their decay widths change with increasing T, have been a source
of intense debate in recent years \cite{pisarski95,pisarski}. In fact models used to examine the 
thermal properties of these vector mesons cover a wide range, starting from 
hadronic models
to calculations involving quarks as fundamental constituent of mesons. If the
phase transition to a QGP phase is primarily one of deconfinement, argues
Pisarski, then this may be modeled by an effective bag constant which decreases
with temperature and the effective mass of the $\rho$ meson, like all hadronic 
bound states, should then decrease with temperature. A gauged sigma model on 
the contrary predicts that masses of $\rho$, $\omega$ and $\phi$ increase 
monotonically with T, while an alternate scenario has been suggested in 
Ref.\cite{br94}.
Recently, Gale and Kapusta,
using an effective model involving pseudoscalar and vector meson, have
showed that the effective masses of vector meson increase with temperature
at leading order of $T^2$ \cite{gale91}, while in QCD sum rules, 
it has been demonstrated 
that with increasing temperature vector and axial vector masses decrease
with the exception of the $\omega$-meson mass which is independent of
temperature \cite{qcd1, qcd2}.
The variation of their decay width with temperature, 
has also been pointed out in Ref.\cite{pisarski95}

In the present letter we focus in particular on the properties of the $\omega$
meson, the shift in its mass and the corresponding change in its 
decay width in a thermal bath. We also calculate the effective $\rho$
meson mass at finite temperature as we use the Gell-Mann Sharp Wagner (GSW)
model of $\rho$ pole dominance to
calculate the matrix element involved in the decay rate for
$\omega\to 3\pi$. In fact, this has a significant 
consequence on the width of $\omega$ meson as shall be explained later.


Vector meson masses, in the present
context are mainly modified by the vacuum excitation of a nucleon-antinucleon
pair at finite temperature. The $\omega$ meson mass,
(as also that of the $\rho$) decreases with increasing temperature
basically due to the nucleon loop and the fact that the effective
nucleon mass (generated by the $\sigma$ meson mean field) is reduced. 
Although because of the reduction in mass of the $\omega$ meson the phase space is 
suppressed, the downward shift in the $\rho$ pole position at finite T 
results in an increase of the matrix element appearing in the 
$\omega$ decay width at finite T. 
Furthermore, the stimulated emission into a pion gas, 
commonly referred to as the Bose Enhancement (BE) would give rise to a
further increase as in the present context the decay is assumed
to take place in a thermal bath and the $\omega$ meson dominantly 
decays into three pions. These factors in fact finally lead to a broader width
for $\omega$ meson in a heat bath compared to what is observed 
at zero temperature. This particular feature, which has not been remarked 
upon earlier, is in contrast to what
has been assumed in Ref.\cite{li96} that the $\omega$ meson decay width
is proportional to the $\omega$ meson mass (which decreases with temperature). 

To calculate the decay width of the $\omega$ meson, as has already been
mentioned, we use 
the GSW model of $\rho$ pole dominance, i.e.
$\omega\to\rho\pi\to 3\pi$ \cite{gell-mann,meis}. 
The $\omega\rho\pi$ interaction Lagrangian can be taken to be
\be
{\cal L}=g_{\omega\rho\pi}
\epsilon_{\mu\nu\alpha\beta}\partial^\mu\omega^\nu\partial^\alpha
\rho^\beta\pi
\ee
The matrix element, therefore, can be calculated  considering the
three channels depending upon the charged state of the intermediate
$\rho$ meson. 
\bea
{\cal M}=2g_{\omega\rho\pi}g_{\rho\pi\pi}m_\omega^\ast
\left[
\frac{\vec\epsilon\cdot (\vec k_+\times\vec k_-)}{(P-k_0)^2-m_\rho^{\ast 2}}
-\frac{\vec\epsilon\cdot (\vec k_+\times\vec k_0)}{(P-k_-)^2-m_\rho^{\ast 2}}
+\frac{\vec\epsilon\cdot (\vec k_-\times\vec k_0)}{(P-k_+)^2-m_\rho^{\ast 2}}
\right ]
\eea
where the asterisk reminds us that instead of the free meson masses the thermal
masses have to be used which are determined from the poles of the dressed
propagator at finite temperature as discussed below. Here P represents 
the momentum of the
$\omega$ meson, the matrix element is written in its rest frame.
i.e. $P=(m_\omega^\ast,0,0,0)$, and $k_\alpha$ with
$\alpha=+,0,-$ denoting the momenta of $\pi^+,\pi^0,\pi^-$. The $\omega\rho\pi$
coupling constant 
can be determined by using the relation 
$g^2_{\omega\rho\pi}= \frac{9g^2_{\rho\pi\pi}}{8\pi^2f_\pi^2}$ 
\cite{gale97}
which, in fact, is consistent with
phenomenological studies, while the $g_{\rho\pi\pi}$ coupling constant is 
determined by fitting the $\rho\to\pi\pi$ decay width. For the coupling 
constants as usual we do not make any finite temperature correction.
The square of the matrix element averaged over the polarization states of
the $\omega$ meson is given by
\be
|{\bar{{\cal M}}}|^2=\frac{4}{3}g_{\omega\rho\pi}^2g_{\rho\pi}^2
m_\omega^{\ast 2} (\vec k_+\times \vec k_-)^2\sum_{\alpha=0,+,-}
\left [ \frac{1}{(P-k_\alpha)^2-m_\rho^{\ast 2}}
\right ]^2
\ee
where the three-momentum conservation condition has been used.

The decay width in the hot pion gas may be written as 
\be
\Gamma_\omega(T)=\frac{1}{8m_\omega^\ast (2\pi)^4}
\int|{{\bar{\cal{M}}}}|^2f_{BE}(E_+,E_-)
dE_+dE_-
\ee
The factor $f_{BE}(E_+,E_-)=\Pi_{\!\!\!\alpha}(1+n(E_\alpha))$ accounts for the BE due to the induced emission
of pions where $n(E_\alpha)$ is the pion distribution function with
$E_0=m^\ast_\omega-E_+-E_-$.

Now, to calculate the decay width integrations have to be performed over
$E_+,E_-$ in the range $E_{+ (max)}=\frac{m_\omega^{\ast 2}-3m_\pi^2}
{2m_\omega^\ast}$ and $E_{+ (min)}=m_\pi$, where the limits of $E_-$ are
$\frac{(m_\omega^\ast - E_+)}{2} \pm \frac{1}{2}
\sqrt{\frac{(E_+^2-m_\pi^2)(m_\omega^{ \ast 2}-2E_+m_\omega^\ast-3m_\pi^2)}
{(m_\omega^{\ast 2}+m_\pi^2-2E_+ m^\ast _\omega)}}$.

In the present study the effective thermal masses of the vector mesons
are generated, as has already been mentioned, by the nucleon-antinucleon
excitation at finite temperature. We need to find the effective masses
of both the $\rho$ and $\omega$ meson. 
The vector meson-nucleon interaction Lagrangian may be written as 
\begin{equation}
{\cal L}_{int} = g_{\alpha} [{\bar{N}} \gamma _\mu \tau^\alpha 
 N - \frac{\kappa _\alpha}{2M}{\bar{N}}
	 \sigma_{\mu\nu}\tau^\alpha N\partial ^\nu]
V^\mu_\alpha 
\end{equation}                                                   
where $V_{\alpha} = \{\omega,\rho\}$,  $\alpha$ running
from 0 to 3, indexes quantities relevant for
$\omega$ when $\alpha = 0$, while 
$\alpha = $ 1 to 3 refers to the $\rho$ meson; $\tau^0 = 1$ and
$\tau^i$ are the isospin Pauli matrices.  The coupling constants
$g_\rho$ and $g_\omega$ as also the ``anomalous" or tensor-coupling parameters
$\kappa_\rho$ and $\kappa_\omega$ may be estimated \cite{sh} 
from the Vector Meson Dominance (VMD) model for nucleon form-factors or 
from the fitting of the nucleon-nucleon interaction data as done 
by the Bonn group \cite{bonn}. In view
of the relatively small value of the iso-scalar magnetic moment of 
the nucleon as compared to the iso-vector part, the tensor coupling
is more important for the $\rho$ than it is for the $\omega$ meson.

The effective vector meson masses, as has already been mentioned , are
determined from the poles of the dressed propagator or the zeros
of the inverse propagator
\be
D^{-1}_{\mu\nu}=D^{-1}_{0\mu\nu} + \Pi_{\mu\nu}
\ee
where the polarization tensor is given by
\begin{equation}
\Pi_{\mu\nu}^{\alpha\beta}={-i}
\int \frac{{d^4}k~}{(2\pi)^4}{\rm Tr}[i\Gamma^\alpha_\mu iS_F(k+q)
{i\bar\Gamma^\beta_\nu} iS_F(k)]
\end{equation}
here $\Gamma$ ${\bar\Gamma}$
represent the appropriate vertex factors and $\alpha$, $\beta$
are the isospin indices.

The nucleon propagator in the real time formalism at finite 
temperature is represented by\cite{dolan}
\begin{equation}
S_F(k)=\frac{k\!\!\!/+M^\ast}{k^2-M^{\ast 2}+i\epsilon}
+\frac{k\!\!\!/+M^\ast}{e^{\beta|u.k|}+1}2\pi i\delta(k^2-M^{\ast 2})
\end{equation}
$u_\mu=(1,0,0,0)$ here defines the thermal bath frame. 
$M^\ast$ is the effective nucleon mass generated by the
sigma meson exchange tadpole at finite temperature.
The separation of the nucleon propagator [eq.(8)] into the T-dependent 
$(S_F^T)$ and the zero temperature contribution $(S_F^0)$
is particularly useful. 

Now clearly, in the polarization tensor there are terms arising from
$S_F^0S_F^0$, $S_F^TS_F^0+S_F^0S_F^T$ and
$S_F^TS_F^T$. The first combination corresponds to the zero temperature
part of the self-energy represented by $Q_{\mu\nu}\Pi^0$, where
$Q_{\mu\nu}=(-g_{\mu\nu} + \frac{q_{\mu}q_{\nu}}{q^2})$ is the
relevant projection tensor.  The $\Pi^0$ integral is 
divergent and is same as that at zero temperature except that 
now it involves 
instead of the free nucleon mass the thermal nucleon mass.
We regularize it using dimensional regularization  and
use the renormalization scheme that the
on-shell vacuum contribution to $\Pi_{\mu\nu}$ 
vanishes at zero temperature.  Expressed mathematically, we
impose the condition 
$\partial^n\Pi^0(q^2)/\partial(q^2) ^n\vert _{M^\ast\rightarrow M,
q^2=m_v^2}=0~ (n=0,1,2...,\infty )$ similar to what one adopts
at finite density ($T = 0$) calculations \cite{sh}
(here and henceforth $\Pi^0(q^2)$ denotes the renormalized quantity).
It may be mentioned that as we are presently dealing with the
fermion loop there is no 
additional source of divergence in the present case at finite temperature. 
Therefore the polarization tensor can be split up into T=0 and $T\neq 0$ part
in the following way
\be
\Pi^0_{\mu\nu}= Q_{\mu\nu}\Pi^0(q^2) + \Pi_{\mu\nu}^{(T)}
\ee
Explicit expressions are given by
\begin{equation}
\Pi_{\mu\nu}^{vv(T)}=\frac{2g_v^2}{\pi ^3}\int\frac{{d^3k}}{E^{\ast}(k)}
\frac{1}{e^{\beta E^\ast(k)}+1}
\frac{{\cal K}_{\mu\nu}q^2-Q_{\mu\nu}(k\cdot q)^2}{q^4-4(k\cdot q)^2}
\end{equation}
\begin{equation}
\Pi_{\mu\nu}^{vt(T)+tv(T)}=\frac{2g_v^2}{\pi ^3}
(\frac{\kappa M^\ast}{4M})2q^4Q_{\mu\nu}\int\frac{{d^3k}}{E^{\ast}(k)}
\frac{1}{e^{\beta E^\ast(k)}+1}
\frac{{1}}{q^4-4(k\cdot q)^2}
\end{equation}
\begin{equation}
\Pi_{\mu\nu}^{tt(T)}=-\frac{2g_v^2}{\pi ^3}
(\frac{\kappa}{4M})^2(4q^4)\int\frac{{d^3k}}{E^{\ast}(k)}
\frac{1}{e^{\beta E^\ast(k)}+1}
\frac{{\cal K}_{\mu\nu}+Q_{\mu\nu}M^{\ast 2}}{q^4-4(k\cdot q)^2}
\end{equation}
\vskip 1 true cm
\noindent where ${\cal K}_{\mu\nu}=(k_\mu-\frac{k.q}{q^2}q_\mu)
(k_\nu-\frac{k.q}{q^2}q_\nu)$ and the superscripts v $\&$ t represent the
vector and tensor coupling contributions.

The effective masses can now be determined from zeroes of the
inverse propagator eq.(6) for $\rho$
and $\omega$ meson which then can be used to determine the $\omega$ meson
decay width in a hot pion gas. Here we take only the nucleon loop as
others like $\rho-\pi$ or $\omega-\pi$ loops for the $\omega$ and
$\rho$ respectively do not contribute much to the
mass shift as compared to the shift brought about by the one considered
here. Actually in the present context the vector meson mass shift is
inextricably related, as mentioned before, to the nucleon mass shift at
finite T because of the $\sigma$ meson mean field. Masses of these hadrons
are reduced mainly 
because the nucleon mass decreases. This is depicted in Fig.(1)

As has already been mentioned we have calculated the vector meson
effective masses at finite temperature using real-time formalism. For the
$\rho$ meson effective mass our findings are consistent with what has been 
obtained using the imaginary-time formalism in Ref\cite{song95}.

We observe from Fig(2) that although the mass of the $\omega$ meson goes
down with temperature thereby suppressing the phase space, 
the decay width shows opposite trend because of the reduction of the $\rho$
meson mass. This is a new observation which indicates that we cannot take the
$\omega$ meson decay width to be
proportional to its mass in contrast to what has been used in Ref\cite{li96}. 
Another important observation is that the Bose enhancement causes
an appreciable increase of the decay width at finite temperature in a pionic
medium. 


In conclusion we observe that $\omega$ meson decay rate in a thermal
bath increases because of an interesting interplay between the phase
space factor (which reduces because of the decrease in $\omega$ meson mass),
the transition matrix element ( which increases because of the 
downward shift of the $\rho$ meson pole) and Bose enhancement (caused
by the stimulated emission of pions at finite temperature). All these
factors are likely to have significant bearings on the dilepton spectra
observed in experiments involving heavy ions.  Such investigations are 
in progress.

We sincerely acknowledge useful discussion with B. Dutta-Roy, B. Sinha,
and J. Alam.

\newpage
\centerline{{\bf Figure captions}}
\begin{enumerate}
\item 
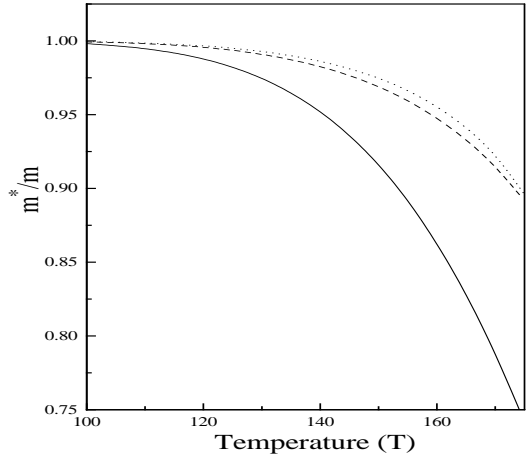
Fig. 1: The variation of the effective nucleon and 
meson
masses with temperature. Solid, dashed and dotted lines represent effective
nucleon, rho meson and omega meson masses respectively.

\item
Fig. 2: 
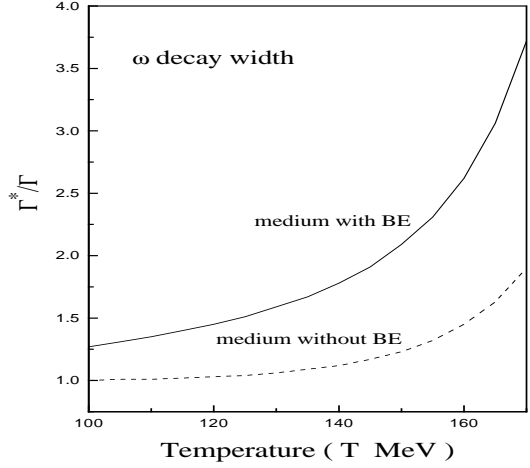
The variation of the decay width of omega meson with
temperature with and without the Bose enhancement factor.
\end{enumerate}
\end{document}